\documentclass[pre,aps,showpacs,nofootinbib,twocolumn]{revtex4}
\usepackage{graphicx}

\begin{document}
\title{Charged rosettes at high and low ionic strengths}
\author{Helmut Schiessel}
\address{Max-Planck-Institut f\"{u}r Polymerforschung, Theory Group, P.O. Box 3148,\\
D-55021 Mainz, Germany}

\begin{abstract}
The complexation between a semiflexible polyelectrolyte and an
oppositely charged macroion leads to a multitude of structures
ranging from tight complexes with the chain wrapped around the
macroion to open multileafed rosette-like complexes. Rosette
structures, expected to occur for short-ranged attractions between
the macroion and the chain, have now also been seen in recent
Monte Carlo simulations with long-range (unscreened) interactions
[Akinchina and Linse {\it Macromolecules} {\bf 2002}, {\it 35},
5183]. The current study provides scaling theories for both cases
and shows that rosette structures are indeed quite robust against
changes in the ionic strength. However, the transition from the
wrapped to the rosette configuration has a dramatically different
characteristics: The short-range case leads to a strongly
discontinuous transition into a rosette with large leaves whereas
the long-range case occurs in a continuous fashion. We provide the
complete diagram of states for both cases.
\end{abstract}
\maketitle

\section{Introduction}

Many biological processes involve the complexation between a
charged macroion and an oppositely charged chain. A prominent
example is the complexation of DNA with histone proteins which is
the basic step of DNA compaction into chromatin in the cells of
animals and plants~\cite {chromatin}. Complexation between
macroions and synthetic polymers is also encountered in many
technological applications as a means to modify macroion solution
behavior; examples are the complexation of polymers with charged
colloidal particles~\cite{ganachaud97,sukhorukov98} and charged
micelles~\cite{mcquigg92}.

There is now a large body of theoretical studies on this set of
problems, most of them have appeared within the last three
years~\cite
{park99,mateescu99,gurovitch99,netz99,kunze00,schiessel00,nguyen01,nguyen01b,
nguyen01c,schiessel01,schiessel01c,kulic02}
(cf. also earlier related studies~\cite{odijk80,pincus84,marky91,vongoeler94}%
). They vary widely with respect to the methods and the level of
approximations used and also with respect to the physical
properties assumed for the chain and the macroion -- the latter
usually being modelled as a charged sphere. Some
studies~\cite{park99,nguyen01c,schiessel01} assume the chain to be
so highly charged that counterion release is the dominant
mechanism for the sphere-chain complexation, whereas most of the
other studies assume weakly charged components attracted via
standard electrostatics. In some cases the ball-chain complexes
interact via a short-range
attraction~\cite{schiessel00,schiessel01c,kulic02,marky91}
corresponding to high ionic strengths~\cite{netz99}. The
investigated systems also vary with respect to the chain
flexibility: semiflexible chains are considered in Refs.~\cite
{park99,netz99,kunze00,schiessel00,schiessel01,schiessel01c,kulic02,marky91}
whereas the other studies assume flexible polymers.
References~\cite {schiessel01c,kulic02} were devoted to the
diffusion (repositioning) of the complexed sphere along the chain.
And, finally, complexes between a chain and several spheres have
been considered~\cite{nguyen01b,schiessel01}. The space available
here does not allow us to give a detailed account of the different
approaches as well as of the phenomenons predicted like
overcharging, conformational symmetries of the adsorbed chain etc.
A
critical survey will be provided in a topical review that is in preparation%
~\cite{schiessel03}.

Sphere-chain complexes have also been investigated in several
computer simulations~\cite
{wallin96,wallin96b,wallin97,wallin98,haronska98,chodanowski01,jonsson01,jonsson02,messina02,messina02b,akinchina02}%
. Wallin and Linse\ studied the effect of chain
flexibility~\cite{wallin96}, line charge density~\cite{wallin96b}
and sphere radius~\cite{wallin97} on the geometry of a complex
between a charged sphere and a polyelectrolyte in a Monte Carlo
simulation that took counterions explicitly into account; they
also considered the case when there are many chains
present~\cite{wallin98}. Chodanowski and
Stoll~\cite{chodanowski01} investigated the adsorption of a
flexible chain on a sphere assuming Debye-H\"{u}ckel interaction.
The case of multisphere adsorption on flexible~\cite{jonsson01}
and semiflexible chains~\cite {jonsson02} was studied by Jonsson
and Linse. Messina, Holm and Kremer~\cite {messina02,messina02b}
demonstrated that in the case of strong electrostatic coupling it
is even possible that a polyelectrolyte chain forms a complex with
a sphere that carries a charge of the {\it same} sign -- a
phenomenon made possible by correlation effects due to
neutralizing counterions. Finally, a recent systematic study by
Akinchina and Linse~\cite{akinchina02} focused again on the role
of chain flexibility on the structure of the sphere-chain complex.

The overall picture emerging from this multitude of theoretical
and simulation studies is still not very clear. Partly this has to
be attributed to the fact that there are many free parameters
determining the properties of the sphere-chain complex, especially
the length of the chain, its linear charge density and persistence
length, the macroion radius and charge and the screening length of
the salt solution. This makes it difficult to develop a theory
that covers the whole range of possible structures. The current
study is an attempt to give a scaling theory that allows an
approximate treatment of the chain-sphere complex over the whole
range of parameters and that especially fills the gaps that were
left open by the existing theories and simulations. It allows to
identify the few independent scaling parameters in this system and
leads to the construction of two-dimensional phase diagrams (one
for short and one for large screening lengths) that cover the
whole range of all the other parameters.

This paper has been induced by a comparison of
Refs.~\cite{schiessel00} and \cite{akinchina02}. In the former
paper we studied the complexation behavior of a semiflexible chain
and a ''sticky'' sphere. In the context of electrostatics this
corresponds to high salt concentrations where the Debye screening
length is much shorter than the sphere radius. In that paper we
calculated the zero-temperature configurations modelling the
polymer by the worm-like chain. We found two typical structures
that occur in this system. If the sphere is sticky enough, i.e.,
if the chain adsorption energy per length is large, the chain
wraps around the sphere -- as long as there is enough surface
available. When the adsorption energy is decreased there is a
point at which the bent chain unwraps in a strongly discontinuous
fashion. The new structure that emerges has multiple point
contacts between the sphere and the chain and large low-curvature
loops connecting them. We called this class of structures the {\it
rosette} configurations.

On the other hand, the latter work, Ref.~\cite{akinchina02},
presented a Monte Carlo study of the complexation of a
semiflexible charged chain with an oppositely charged ball that
carries the same absolute charge as the chain (isoelectric
complex). No small ions were present so that the charged monomers
were attracted to the sphere via an unscreened long-range
$1/r$-interaction. The authors simulated systems with different
persistence lengths, linear charge densities of the chain and ball
radii. Depending on the choice of parameters they encountered a
multitude of structures -- ranging from collapsed structures with
a ''tennisball seam pattern'' or solenoid arrangement of the
wrapped chain~\cite{kunze00} to open multi-leafed structures very
much resembling the ones found in Ref.~\cite{schiessel00}. Their
rosette structures occur for stiffer chains on smaller spheres.

That work demonstrates that the rosette structure is quite robust
with respect to the range of interaction. At the same time it also
hints towards major differences between these two cases: The
short-ranged case predicts clearly a strongly discontinuous
unwrapping transition into large-leafed rosettes when the
adsorption energy is decreased or the chain stiffness is
increased~\cite{schiessel00}. In the
simulation~\cite{akinchina02}, however, the rosette evolved
continuously with increasing chain stiffness from a tightly
wrapped state via slightly more open structures with many small
loops to large-leafed rosettes.

How can one reconcile these findings? For that purpose we
reconsider in Section II the chain-sphere complexation for the
case of short-ranged attraction~\cite{schiessel00}, formulating
now the interaction in the language of strongly screened
electrostatics instead of some short-ranged stickiness. Then, in
Section III, we give the scaling description in the case of weak
screening. By focusing on the unwrapping transition we contrast
the two cases in Section IV. In the final section we compare our
findings to computer simulations and experimental observations on
the DNA-histone complex.

\section{Charged rosettes at high ionic strength}

Consider a polymer chain of length $L$, radius $r$ and persistence length $%
l_{P}$. The chain carries negative charges on its backbone -- a distance $b$
apart -- which leads to a linear charge density $-e/b$. The macroion is
modelled as a sphere of radius $R$ that carries a positive charge $Z$. The
reduced electrostatic potential is assumed to be smaller than unity
everywhere so that the electrostatic interaction can be described by
standard Debye-H\"{u}ckel theory: Two elementary charges at distance $r$
interact with the potential $e\phi /kT=\pm l_{B}e^{-\kappa r}/r$ with $%
l_{B}=e^{2}/\varepsilon k_{B}T$ ($\varepsilon $: dielectric
constant of the solvent, $k_{B}T$: thermal energy), the Bjerrum
length, and $\kappa ^{-1}=\left( 8\pi l_{B}c_{s}\right) ^{-1/2}$,
the Debye screening length for
monovalent salt of concentration $c_{s}$. $l_{B}$ is of the order of $7$\AA\ %
in water at room temperature.

In this section we assume strong screening such that the screening
length is much smaller than the radius of the ball, $\kappa R\gg
1$. We will assume throughout the chain to be so thin that $r$ is
the smallest length scale in our system, especially that always
$\kappa r\ll 1$. The adsorption energy per length can then be
estimated from the Debye-H\"{u}ckel electrostatic potential close
to the surface of the sphere which mimics for $\kappa R\gg 1$ that
of a plane with a surface charge density $\sigma =Z/\left( 4\pi
R^{2}\right) $. One has then $e\phi /kT=l_{B}Z\kappa
^{-1}e^{-\kappa z}/R^{2}$ where $z$ is the distance from the
surface (cf., for instance, Ref.~\cite{netz99}). This leads to the
following adsorption energy per length (in units of $k_{B}T$)
\begin{equation}
\lambda \simeq \frac{l_{B}Z}{b\kappa R^{2}}  \label{lamda}
\end{equation}
Now the chain will only wrap on the sphere when $\lambda $ is
large enough, namely so large that it exceeds $l_{P}/2R^{2}$, the
bending energy per length~\cite{harries66}. The (free) energy of
the wrapped chain-sphere complex is then given by
\begin{equation}
\frac{F_{wrap}}{k_{B}T}\simeq \frac{l_{P}L}{R^{2}}-\lambda L=\left( \frac{%
l_{P}}{R^{2}}-\lambda \right) L  \label{fwrap}
\end{equation}
This leads to the prediction of an unwrapping transition at
$\lambda =l_{P}/R^{2}$ when the chain stiffness
\begin{equation}
l_{P}\simeq \frac{l_{B}Z}{b\kappa }  \label{unwrap}
\end{equation}
is reached. Since the wrapping path of the chain on the sphere can
be quite intricate (cf. Ref.~\cite{kunze00}) the local radius of
curvature is not always precisely $R$ (but of the order $R$). Thus
we dropped from Eq.~\ref{fwrap} on all numerical prefactors and
will also do so in the rest of the paper (for similar reasons).
Such an unwrapping transition has been first discussed by Marky
and Manning~\cite{marky91} using some unspecified short-range
attraction and by Netz and Joanny~\cite{netz99} for the
electrostatic case (in fact, their Eq.~(35) coincides with
Eq.~\ref{unwrap}).

For larger persistence lengths than the one given in Eq.
\ref{unwrap} the chain has to completely unwrap from the sphere.
This led the authors of Ref.~\cite{marky91} to the prediction of
an ''all-or-nothing'' picture: either the chain is wrapped or it
is unwrapped -- with a single adsorption point or even completely
desorbed. We showed however in a later study that the chain can
lower its energy considerably by having multiple point contacts to
the
sphere which leads to the rosette structures~\cite{schiessel00}. Let us call $%
\mu $ the energy per contact point (in units of $k_{B}T$). Its value
\begin{equation}
\mu \simeq \lambda \sqrt{R\kappa ^{-1}}\simeq \frac{l_{B}Z}{b\left( \kappa
R\right) ^{3/2}}  \label{mu}
\end{equation}
follows from the length $\sqrt{R\kappa ^{-1}}$ of a chain portion
around a point contact that is located within the distance $\kappa
^{-1}$ from the sphere. The free energy of an $N$-leafed rosette
has then approximately the following form
\begin{equation}
\frac{F_{rosette}}{k_{B}T}\simeq \frac{l_{P}}{L}N^{2}-\mu \left( N+1\right)
\label{leaves1}
\end{equation}
The first term describes the energy required to bend the $N$ chain
pieces between the point contacts, each of length $L/N$, into
leaves with typical curvature $L/N$, the second term account for
the $N+1$ point contacts. We do not account here for the entropy
of the chain configurations that can be safely neglected for large
point contact energies $\mu \gg 1$ (see below). The optimal number
of leaves follows from the minimization of the (free) energy to be
\begin{equation}
N^{*}\simeq \mu \frac{L}{l_{P}}  \label{nstar}
\end{equation}
which leads to leaves of size
\begin{equation}
l_{leaf}\simeq \frac{L}{N^{*}}\simeq \frac{l_{P}}{\mu }  \label{lleaf}
\end{equation}
As already pointed out in Ref.~\cite{schiessel00} each leaf
minimizes its bending energy by assuming the shape of a so-called
Yamakawa-Stockmayer (YS) loop, namely a lemniscate-shaped loop
with an 81-degree apex angle~\cite {yamakawa72} (cf. also
Ref.~\cite{kulic02} for an approximate derivation of the YS loop).
As a cautionary remark we mention that the bending energy of YS
loops leads to the numerical prefactor 14.04 for the first term in
Eq.~\ref{leaves1}~\cite {yamakawa72} -- clearly demonstrating that
one should not extract from this theory any numbers but just
scaling relations. The leaves show negligible shape fluctuations
for $l_{leaf} \ll l_P$, i.e. for $\mu \gg 1$. Then also $N^{*}$,
Eq.~\ref{nstar}, is essentially fixed and the rosette has a
well-defined leaf number -- as implicitely assumed here.

\begin{figure}
\includegraphics*[width=8cm]{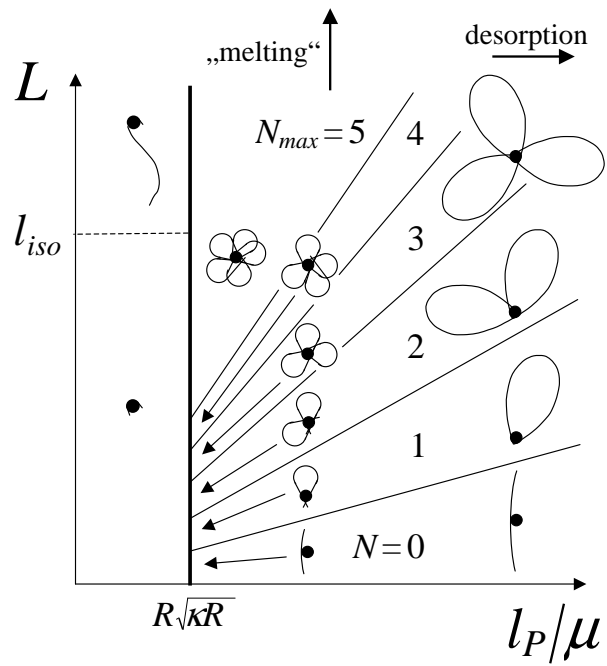}
\caption{The sphere chain complex at high ionic strength, $\kappa
R\gg 1$. Depicted is the "phase" diagram as a function of the
total length $L$ of the chain and of its persistence length $l_P$
divided by the point contact energy $\mu$. The thick vertical bar
denotes the discontinuous unwrapping transition. To the left are
the wrapped complexes, to the right the rosettes.}
\end{figure}

In Fig.~1 we summarize our results in a ''phase diagram'' of total
chain length $L$ vs. $l_{P}/\mu $ (which is in the case of
rosettes nothing but the
leaf-size, cf. Eq.~\ref{lleaf}). For small values of $%
l_{P}/\mu $ we have wrapped structures, for large values rosette
conformations. The unwrapping transition occurs at $l_{P}/\mu \simeq R\sqrt{%
\kappa R}$ (cf. Eqs.~\ref{unwrap} and \ref{mu}) in a strongly
discontinuous fashion indicated by a thick vertical line. The
other lines indicate transitions between different ground states,
namely rosettes with different numbers of leaves.

Additional features arise from the fact that the sphere surface is
finite. As we will show now this can lead to the formation of
tails for the case of wrapped structures and to the melting of
rosettes. Consider first a wrapped chain configuration, i.e.,
assume some fixed value for $l_{P}/\mu <R\sqrt{\kappa R} $, and
increase $L$, i.e., go along a vertical line in the diagram,
Fig.~1. It is clear that at a certain point the wrapping has to
stop. Locally the wrapped chain forms a nearly planar lamellar
phase with a distance $d$ between neighboring chain sections. This
distance follows from the competition between the chain-sphere
attraction and the repulsion between neighboring chain
segments~\cite{netz99,netz99b,schiessel01b}
\begin{equation}
d\simeq \frac{1}{b\sigma }\simeq \frac{R^{2}}{bZ}  \label{dist}
\end{equation}
This leads to a wrapping length $l\simeq R^{2}/d\simeq bZ\simeq
l_{iso}$ which is just the isoelectric wrapping length, the length
at which the wrapped chain portion just compensates the sphere
charge. Chains that are shorter than this length, $L<l_{iso}$,
will be completely wrapped, chains
that are longer, $L>l_{iso}$, will have their extra length $%
L-l_{iso}$ dangling from the complex in the form of one or two
tails. The tail formation at $L=l_{iso}$ has been indicated in
Fig.~1 by a horizontal dashed line. We also note that if $d>\kappa
^{-1}$ in Eq.~\ref{dist}, i.e., if $Z<R^{2}\kappa /b$ (large
sphere), the lamellar spacing is reduced to the value $\kappa
^{-1}$ as pointed out by Netz and Joanny~\cite{netz99}.

Now let us consider the rosettes. Choose some arbitrary value for $l_{P}/\mu
>R\sqrt{\kappa R}$ and increase $L$. Then each time when the chain length is
increased by an amount $l_{P}/\mu $ an additional leaf is formed.
However, packing constraints imply an upper limit for the number
of loops of the order $N_{\max }\approx l_{iso}/\sqrt{R\kappa
^{-1}}$; this is the maximal number of contacts, each excluding an
area $\approx d\sqrt{R\kappa ^{-1}}$, that can be closely packed
on the surface of the sphere. For a
rosette with the maximal number of leaves (in Fig.~1 we chose arbitrarily $%
N_{\max }=5$) an increase in $L$ leads to an increase of the leaf
size. $l_{leaf}$ reaches the persistence length when $L\simeq
N_{\max }l_{P}$. At that point melting of the rosette takes place,
i.e., for larger values of $L$ the leaf size distribution is
heterogeneous, the YS loops show shape fluctuations and also the
tails start to grow. Entropic effects are then important. The
thermodynamics of this melting process is quite intricate and can
be calculated by mapping this problem on an exactly solvable
one-dimensional many body problem~\cite{schiessel00,schiessel03}.
In the current study we will not discuss the rosette melting
further and merely indicate it in Fig.~1 by an arrow pointing
towards the direction of the $L$-axis. Also shown in this figure
is the direction where the chain desorption is to be expected. It
should occur when the complexation free energy of the rosette,
Eq.~\ref
{leaves1}, becomes smaller than $k_{B}T$, i.e. for $l_{P}/\mu >\sqrt{l_{P}L}$%
.

\section{The rosette state at low ionic strength}

The rosette configurations discussed in the last section occur for
chains that are so stiff that wrapping would be too costly. It
allows a small fraction of the chain to be close to the sphere in
the form of point contacts. The majority of the monomers resides
in the loops that do not ''feel'' the presence of the sphere but
are needed to connect the point contacts via low curvature
sections. At first sight one might thus expect the rosette
configurations to be a special feature for chain-sphere complexes
with a short ranged mutual attraction.

That this is not true has already been pointed out in the
introduction. As observed in the simulations by Akinchina and
Linse~\cite{akinchina02} rosettes are in fact quite robust and
occur also in systems with a much larger range of interaction. We
shall give now a scaling theory that provides us with a diagram of
states for the chain-sphere complex for the
case of weak screening, $\kappa R\ll 1$. We will first treat the case of short chains $%
L=bN\leq bZ=l_{iso}$ where the chain charge is smaller than (or
equals) the sphere charge. The free energy of the rosette with $N$
leaves is then approximately given by
\begin{equation}
\frac{F_{rosette}}{k_{B}T}\simeq \frac{l_{P}}{L}N^{2}-\frac{l_{B}Z}{b}N
\label{rosette1}
\end{equation}
The first term is the bending energy of $N$ leaves as already
discussed after Eq.~\ref{leaves1}. The second term accounts for
the attraction between
the ball charge $Z$ and the chain charge $L/b$ over the typical distance $%
L/N $. All other contributions to the electrostatics like the
monomer-monomer repulsion are smaller and neglected in
Eq.~\ref{rosette1}. Remarkably this free energy has the same form
as the free energy~\ref{leaves1} of the rosette for short-ranged
interaction. We just have to identify the point contact energy
$\mu $ (for the strong screening case) with the single leaf-sphere
interaction
\begin{equation}
\mu \simeq \frac{l_{B}Z}{b}  \label{rosette2}
\end{equation}
for the unscreened case. The optimal leaf number is thus again
given by Eq. \ref{nstar} and the leaf size by Eq.~\ref{lleaf} but
now with $\mu $ given by Eq.~\ref{rosette2}. Note that these
scaling results remain even true if the leaves grow so large that
their outer sections do not interact with the sphere due to
screening, $\kappa l_{leaf}>1$. This is so because the
electrostatic rosette-sphere attraction still scales like
$-l_{B}ZN/b$: a fraction $\kappa ^{-1}/\left( L/N\right) $ (i.e.
$\kappa ^{-1}/b$ charges)
of each of the $N$ leaves interacts with the sphere at a typical distance $%
\kappa ^{-1}$.

The rosette state competes with the wrapped state. We expect that the
rosette state is {\it continuously} transformed into the wrapped state when $%
l_{leaf}\simeq R$; then the leaves become so small that they touch with
their contour the surface of the sphere. Indeed, by setting $N=L/R$ in Eq.
\ref{rosette1} one finds
\begin{equation}
\frac{F_{wrap}}{k_{B}T}\simeq \frac{l_{P}L}{R^{2}}-\frac{l_{B}ZL}{bR}=\left(
\frac{l_{P}}{R^{2}}-\frac{l_{B}Z}{bR}\right) L  \label{rosette4}
\end{equation}
which can be considered as the free energy of the wrapped state:
The first term is the bending energy required to wrap the chain
around the sphere and the second accounts for the electrostatic
attraction between the wrapped chain and the sphere. All other
electrostatic contributions (as discussed in detail by Nguyen and
Shklovskii~\cite{nguyen01}) are less important and do not occur on
this level of approximation. On the right hand side of Eq.
\ref{rosette4} we arranged the terms in such way that one can
deduce directly an unwrapping transition, namely at
$l_{P}/R^{2}=l_{B}Z/bR$ which can be rewritten as
\begin{equation}
l_{P}\simeq \mu R\simeq \frac{l_{B}ZR}{b}  \label{rosette5}
\end{equation}
Comparing Eqs.~\ref{fwrap} and \ref{rosette4} one might expect
that the chain unwraps in a strongly discontinuous fashion as
discussed in the previous section. However, this ''unwrapping
point'' corresponds just to the point $l_{leaf}\simeq R$ when
small loops start to grow on the sphere; we thus expect the
transition to be continuous as pointed out before
Eq.~\ref{rosette4}. That the unwrapping transition occurs rather
smooth at low ionic strength and in a strongly discontinuous way
at high ionic strength has been predicted by Netz and
Joanny~\cite{netz99}; however, in that study the authors did not
allow for rosette structures.

To complete the picture we finally have to consider chains that are longer
than the isoelectric length, $L>l_{iso}$. We need then at least three terms
to capture the essential physics of the rosette state:
\begin{equation}
\frac{F_{rosette}}{k_{B}T}\simeq \frac{l_{P}}{l}N^{2}-\frac{l_{B}Z}{b}N+%
\frac{l_{B}l}{b^{2}}N  \label{rosette6}
\end{equation}
Here we ''allow'' the monomers to distribute between the rosette of length $%
l $ and a tail of length $L-l$. The first two terms in
Eq.~\ref{rosette6} are then as above, Eq.~\ref{rosette1}, the last
term describes the repulsion of the monomers that constitute the
rosette. The contributions from the tail can be neglected.
Minimization with respect to $l$ leads to the optimal rosette
length
\begin{equation}
l^{*}\simeq b\sqrt{\frac{l_{P}N}{l_{B}}}  \label{rosette7}
\end{equation}
The free energy \ref{rosette6} with the optimal length $l^{*}$,
Eq.~\ref {rosette7}, is minimized for the following number of
leaves:
\begin{equation}
N^{*}=\mu \frac{l_{iso}}{l_{P}}  \label{rosette8}
\end{equation}
Hence -- on this level of approximation -- $l^{*}\simeq b\sqrt{%
l_{P}N^{*}/l_{B}}\simeq l_{iso}$, i.e., the rosette monomers just
compensate the central ball charge; the rest of the monomers
extends away from the rosette in one or two tails of total length
$L-l_{iso}$. Each leaf is of size
\begin{equation}
l_{leaf}\simeq \frac{l_{P}}{\mu }  \label{rosette9}
\end{equation}
The rosette disappears for $l_{iso}/N^{*}=R$, i.e., when
Eq.~\ref{rosette5} is fulfilled. It is then replaced by a wrapped
structure of length $l_{iso}$ and one or two tails having the
total length $L-l_{iso}$.

\begin{figure}
\includegraphics*[width=8cm]{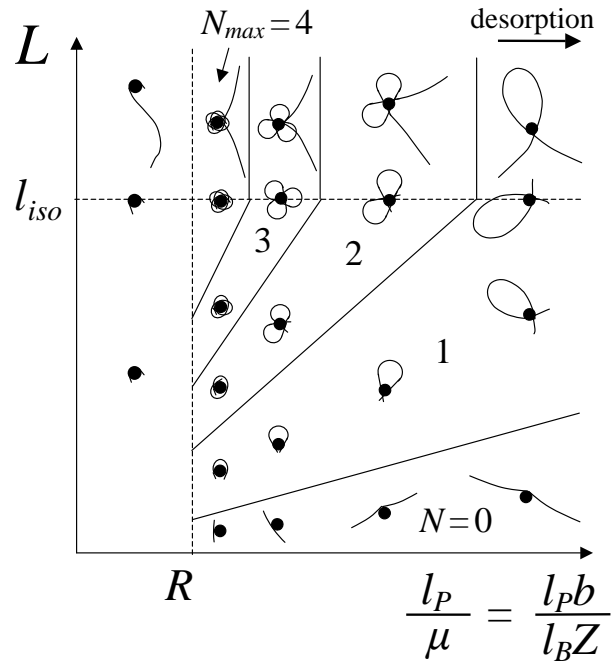}
\caption{Diagram of states for the case of low ionic strength,
$\kappa R\ll 1$. The axes are chosen similar to Fig.~1 with $\mu$
now being the leaf-sphere attraction in the rosette structures.
The unwrapping is here rather smooth (dashed vertical line).
Complexes with long chains, $L>l_{iso}$ show
 tails.}
\end{figure}

In Fig.~2 we depict the complete diagram of the sphere-stiff chain
complexes to be expected in the unscreened case. We again plot $L$
vs. $l_{P}/\mu $, the leaf size. When one starts in this diagram
at a large value of $l_{P}/\mu $ and goes towards smaller values
of (with some arbitrarily fixed value $L<l_{iso}$) then all leaves
shrink and more and more leaves can form. At $l_{P}/\mu =R$ the
maximal number of leaves (for that given value of $L$) is reached
and at the same time the leaves disappear simultaneously in a
continuous fashion. For $l_{P}/\mu <R$ the chain wraps around the
sphere. For long chains, $L>l_{iso}$, the excess charges are
accommodated in tails and all rosettes have the same length
$l_{iso}$. The borderlines between different rosette ground states
are then independent of the total length of the chain and thus
appear as vertical lines.

Desorption occurs when the free energies, Eqs.~\ref{rosette1} and
\ref
{rosette6}, equal the thermal energy $k_{B}T$. This point is reached when $%
l_{P}/\mu \simeq \sqrt{l_{P}L}$ for short chain, $L\leq l_{iso}$, and when $%
l_{P}/\mu \simeq \sqrt{l_{P}l_{iso}}$ for long chains,
$L>l_{iso}$. We indicate in Fig.~2 the direction where desorption
occurs by an arrow.

\section{Unwrapping at high and low ionic strength}

We take now a closer look at the unwrapping transition,
contrasting the short- and the long-range case. The former case,
$\kappa R\gg 1$, is depicted in Fig.~3(a). As discussed in Section
II we expect the unwrapping transition to occur at $\lambda \simeq
l_{P}/R^{2}$ which leads to Eq.~\ref {unwrap}. At this point the
structure jumps in a strongly discontinuous fashion into a {\it
large}-leafed rosette with leaves of size $l_{P}/\mu $, cf.
Eq.~\ref{lleaf}. Combining Eqs.~\ref{unwrap}, \ref{mu} and
\ref{lleaf} we find indeed
\begin{equation}
l_{leaf}\simeq R\sqrt{\kappa R}\gg R  \label{critleaf}
\end{equation}
As discussed in Section II each leaf has the shape of a
YS-loop~\cite {yamakawa72} with an 81-degrees apex angle. This
means that neighboring
leaves have a relative orientation $\theta =180{{}^{\circ }}-81{{}^{\circ }}%
=99{{}^{\circ }}$, cf. Fig.~3(a). In addition, the leaves have to
be slightly twisted (like propeller blades) to account for the
mutual excluded volume.

\begin{figure}
\includegraphics*[width=8cm]{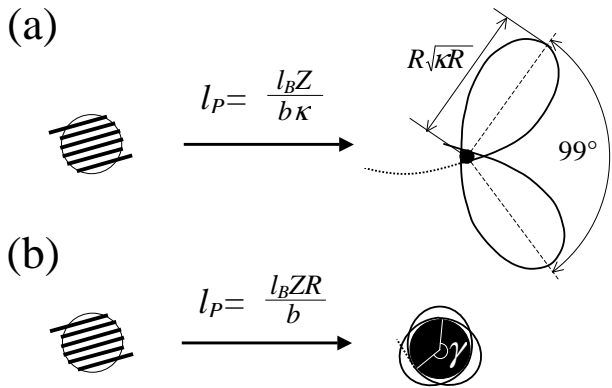}
\caption{The unwrapping transition (a) at high ionic strength,
$\kappa R\gg 1$ and (b) at low ionic strength, $\kappa R\ll 1$. In
the former case the transition is discontinuous and large
Yamakawa-Stockmayer loops are formed, in the ladder case the
rosette structures evolves continuously.}
\end{figure}

The unwrapping at low ionic strength is depicted in Fig.~3(b) and
goes as follows (we discuss here the case $L\leq l_{iso}$; in the
opposite case one
has just to replace $L$ by $l_{iso}$). When the chain becomes so stiff that $%
l_{P}/R^{2}>l_{B}Z/\left( bR\right) $ the wrapped state is not stable
anymore (cf. Eq.~\ref{rosette4}). At that point many small leaves ($%
N^{*}=L/R $ ones) form simultaneously in a continuous fashion.
Their size scales as $l_{leaf}\simeq R$, the precise prefactor
being not accessible to our scaling argument. The number of
windings around the ball scales as $L/R$ (but note that the
geometry of the wrapped path can be actually quite intricate as
demonstrated in Ref.~\cite{kunze00}) and the typical opening angle
$\gamma $ of each loop at the point of its formation scales as
$\left( L/R\right) /N^{*}\approx 1$, again with an unknown
numerical value. A multi-leafed configuration slightly above the
unwrapping point is depicted schematically on the right hand side
of Fig.~3(b).

Additional insight can be gained by generalizing the attractive force
between a given chain charge and the sphere by a power law $-AZ/r^{\alpha }$
with an arbitrary exponent $\alpha >0$. An integer value $\alpha =D-2$ with $%
D=3,4,...$ can be interpreted as a charged chain that adsorbs on
an oppositely charged $D$-dimensional ball in $D$ dimensions. The
electrostatic term for the rosette in Eq. \ref{rosette1} takes
then the form $-AZN^{\alpha }/\left( bL^{\alpha
-1}\right) $ and the one for the wrapped state scales as $-AZL/bR^{\alpha }$%
. Unwrapping takes place at $l_{P}^{*}\simeq AZ/\left( bR^{\alpha -2}\right)
$. At this critical value the free energy of the rosette $F_{rosette}\left(
l_{P}=l_{P}^{*}\right) $ has (as a function of $N$) a minimum at $%
N^{*}\simeq L/R$ for $\alpha <2$ ($D<4$), suggesting a rather
smooth unwrapping transition similar to the one depicted in
Fig.~3(a). This minimum turns into a maximum at $\alpha =2$
($D=4$). For larger values of $\alpha $ we find $N^{*}=0$, i.e.,
the unwrapping transition is strongly discontinuous, similar to
the short-ranged case discussed in the previous section.

Finally we note that in the case of very highly charged chains and
spheres the Debye-H\"{u}ckel approximation breaks down and
nonlinear phenomenons like counterion condensation become
important~\cite{oosawa71,manning78}. The dominant contribution to
the complexation energy is then the release of counterions that
were condensed on the sphere and on the chain before complexation.
This is a short-ranged interaction and consequently the unwrapping
transition is expected to be discontinuous -- even at low ionic
strength. Following Ref.~\cite{schiessel01} the wrapped state can
be written as
\begin{equation}
\frac{F_{wrap}}{k_{B}T}\simeq \left( \frac{l_{P}}{R^{2}}-\frac{\Omega }{b}%
\right) L  \label{fwrap2}
\end{equation}
where $\Omega $ is the free energy gain (in units of $k_{B}T$) per adsorbed
chain charge which follows from the entropy gain due to counterion release ($%
\Omega $ is a number of order one; for details
cf.~\cite{schiessel01}). Unwrapping into the rosette takes place
at $l_{P}\simeq \Omega R^{2}/b$. The point contact energy is of
the order $\left( \Omega /b\right) \sqrt{R\lambda _{GC}}$ where
the so-called Guoy-Chapman length $\lambda _{GC}\simeq 1/\left(
\sigma l_{B}\right) $ is the thickness of the layer of condensed
counterions around the sphere (which is always much smaller than
$R$ when there is strong counterion condensation). The leaf size
at the unwrapping point is then given by
\begin{equation}
l_{leaf}\simeq \sqrt{\frac{R}{\lambda _{GC}}}R\gg R  \label{lleaf2}
\end{equation}
which indeed indicates a strongly discontinuous unwrapping transition.

\section{Discussion and Conclusion}

We compare now our results with Monte Carlo simulations by
Akinchina and Linse~\cite{akinchina02} and then with the behavior
of a biological chain-macroion complex, the
nucleosome~\cite{chromatin}. The simulated
systems~\cite{akinchina02} were always at the isoelectric point,
i.e., the chain length was given by $L=bZ=l_{iso}$. Furthermore,
the charges in the simulation interacted via a non-screened
Colombic $1/r$ potential. The simulation results have thus to be
compared with Fig.~2. Four systems have been considered, each
having having a fixed set of parameters $b$, $Z$ and $R$ but with
7 different values of $l_{P}$. This means that for each case the
systems were located on the dashed horizontal line at $L=l_{iso}$
in Fig.~2.

In one system (called system II in Ref.~\cite{akinchina02}) the
continuous development from a wrapped to the rosette
configurations has been seen most clearly. Example configurations
are shown in Fig.~1, system II in that paper (that these are
representative can be seen by inspecting the adsorption
probability of monomers as a function of the monomer index, cf.
Fig.~3 in~\cite{akinchina02}). For $l_{P}=7$\AA\ the chain is wrapped, at $l_{P}=60$%
\AA\ there is already a slight indication of very small loops
($N=4$ or $5$, cf. the small oscillations in Fig.~3, systems II,
open squares). The next system depicted has already a much stiffer
chain, $l_{P}=250$\AA , and shows very clearly three leaves, then
two leaves at $l_{P}=500$\AA\ and one leaf for the stiffest chain,
$l_{P}=1000$\AA . In Fig.~2 we have chosen the parameters such
that $N_{\max }\simeq l_{iso}/R$ equals 4 so that this corresponds
roughly to system II in~\cite{akinchina02}. To compare with the
simulations we have to follow the $L=l_{iso}$-line in Fig.~2: One
starts with wrapped structures for $l_{P}/\mu <R$ and finds then
the continuous evolution of rosettes when the line $l_{P}/\mu =R$
is crossed. The leaves grow at the expense of their number (first
4, then 3, 2 leaves), just as it has been observed in the
simulations.

In another system (system I) all parameters were kept the same
except the sphere radius that was now twice as large. This system
shows 3 small loops
when $l_{P}=250$\AA\ is reached and one loops for $l_{P}=500$\AA\ and $1000$%
\AA . In our ''phase diagram'' the transformation $R\rightarrow
2R$ means that the unwrapping transition line moves to the right
and consequently the maximal number of loops goes down by a factor
of two: $N_{\max }\rightarrow N_{\max }/2$ which is in
satisfactory agreement with the finding in~\cite {akinchina02}.
Going from our reference system (II) to another case (system IV)
means to go to a higher line charge density of the chain, i.e. the
transformation $b\rightarrow b/2$ (and thus $l_{iso}\rightarrow
l_{iso}/2$). In that case the horizontal line $L=l_{iso}$ has to
be moved to smaller value which leads again to $N_{\max
}\rightarrow N_{\max }/2$. The complex in Ref.~\cite{akinchina02}
shows an unwrapping only for $l_{P}>500$\AA\ and has already a
well pronounced loop at $l_{P}=1000$\AA . Apparently, the wrapped
state is more stable in this system than in system I -- and the
scaling argument predicts indeed the critical persistence length
to scale as $1/b$, cf. Eq.~\ref{rosette5}. The last system (III)
finally combines a large sphere, $R\rightarrow 2R$, with a short
chain, $b\rightarrow b/2$. In that case no indication of loops has
been observed. Even at the largest persistence length the chain is
touching the sphere over some finite section (similarly to the
geometry discussed by Netz and Joanny~\cite{netz99}). In fact, we
predict $N_{\max }\rightarrow N_{\max }/4$ by going from the
reference system (II) to this case. Altogether, our scaling
approach is in quite satisfactory agreement with the observations
made in the Monte Carlo simulations~\cite{akinchina02}.
Interesting would be to test our predictions also for chains that
are shorter or longer than the isoelectric length. Corresponding
simulations are already on the way~\cite{anna}

Now we compare our scaling results with an experimental system,
the nucleosome~\cite{chromatin}. The wrapping of DNA around
protein spools, octamers of so-called histones, is the basic step
in the condensation of DNA into the dense chromatin complex inside
the nuclei of plant and animal cells. Each repeating unit of this
complex, wrapped DNA, histone octamer and a stretch of linker DNA
(connecting to the next such protein spool) is called nucleosome.
When the linker is digested away the remaining complex, the
so-called nucleosome core particle, consists of 147 basepairs
($\approx 500$\AA ) DNA wrapped in 1-and-3/4 left-handed
superhelical turns around the octamer. The core particle which has
a radius of $\sim 50$\AA\ and a height of $\sim 60$\AA\ is
documented in great detail on the basis of high-resolution x-ray
analyses~\cite{luger97}. The histone octamer contains 220 basic
side chains (arginine and lysine)~\cite{khrapunov97}. From these
are about 103 located in flexible histone tails that dangle of the
core particle and/or are complexed with the wrapped
DNA~\cite{mangenot02}. The rest, 117 residues, are in the globular
part of the octamer, of which 31 are exposed to the solvent, the
rest being involved in intra- and interprotein ionic interactions.
On the other hand, one has 147 bps of DNA wrapped around the
octamer, each contributing two phosphate groups. Hence there are
294 negative charges from the DNA versus 220 positive charges of
the octamer, i.e., the nucleosomal complex is overcharged by the
DNA.

Yager, McMurray and van Holde~\cite{yager89} characterized the
stability of the nucleosome core particle as a function of the
salt concentration (NaCl). Using a variety of experimental methods
(e.g. gel electrophoresis) they arrived at the following main
conclusions: The core particle is stable for ionic strengths
ranging from $2mM$ to $750mM$ (this includes physiological
relevant salt concentrations around $100mM$). For slighly higher
salt concentrations the DNA is partially dissociated; an
equilibrium between histones, free DNA and core particles is
observed. At salt concentrations beyond $1.5M$ the core particle
is completely dissociated into histone
oligomers and free DNA. On the other end, for very low salt concentration $%
\lesssim 1mM$ one finds an ''expanded'' state of the nucleosome.
Khrapunov et al.~\cite{khrapunov97} came via fluorescence
measurements to similar conclusions: For ionic strengths between
$5$ and $600mM$ the core particle is intact. At larger ionic
strength ($\approx 1.2M$) the terminal regions of the DNA unwrap
and two histone dimers are dissociated and at a even larger value
($\approx 1.5M$) the remaining tetramer leaves the DNA. Finally,
at low salt concentrations one encounters an open state: the
dimers break their contact with the tetramer and the DNA termini
unwrap.

The key features of the behavior of core particle DNA (neglecting
the substructure of the octamer) were recovered in a numerical
study by Kunze and Netz~\cite {kunze00}. They considered the
complexation of a charged, semiflexible chain with an oppositely
charged sphere, interacting via a screened Debye-H\"{u}ckel
potential. The optimal DNA configuration was derived numerically
from the minimization of the energy. For a reasonable set of
parameters (comparable to the one of the core particle) they found
for vanishing ionic strength ($\kappa ^{-1}\rightarrow \infty $)
an open, planar configuration where only a small fraction of the
chain is wrapped whereas the two tails (of equal length) are
extended into roughly opposite directions. This is reminiscent of
the open structures reported in the experimental
studies~\cite{khrapunov97,yager89}. Upon addition of salt the
structure goes from a two- to a one-tail configuration and the
chain begins to wrap more and more. Already well below
physiological ionic strengths the chain is almost completely
wrapped. The chain stays in this wrapped state up to very high
salt concentrations. Only then the chain unwraps in a
discontinuous fashion when the chain-sphere attraction is
sufficiently screened. Also these features of the complex (a
wrapped compact state in a wide range around physiological
conditions and unwrapping at high salt content) reflect the
experimental finding in Refs.~\cite{khrapunov97} and
\cite{yager89}.

How are these findings related to our scaling study? Let us start
at physiological salt concentrations around $100mM$. Then $\kappa
^{-1}\simeq 10 $\AA\ is much smaller than the particle size
$\approx 100$\AA\ and we have the short-range case. Since $r\simeq
10$\AA , i.e., $\kappa r\approx 1$ for physiological conditions,
and since the binding sites between DNA and the histones are quite
specific~\cite{luger97}, the estimation of $\lambda $ in Eq.
\ref{lamda} is not reliable. $\lambda $ can be derived instead
experimentally from competitive protein binding to nucleosomal
DNA~\cite {polach95} to be of the order $\left( 1/5\right)
k_{B}T/$\AA\ which is roughly 10 times smaller than what
Eq.~\ref{lamda} would predict (but note also that contact is made
mainly between DNA minor grooves which are 10 basepairs apart!).
This short-range attraction is largely balanced by the strong
bending contribution ($l_{P}=500$\AA\ for DNA and hence
$l_{P}/\left( 2R^{2}\right) \simeq 1/6-1/7k_{B}T/$\AA ). In any
case, Eq. \ref{unwrap} predicts an unwrapping for sufficiently
small values of $\kappa ^{-1}$ but the numbers are not reliable.
The more interesting case is that of the core particle at low
ionic strength around $1mM$ ($\kappa ^{-1}\simeq 100$\AA ). At
that salt concentration $2\kappa R\approx 1$ and we enter the
long-range case. Apparently, the completely wrapped configuration
is not stable at this point anymore. This is not surprising since
the nucleosomal DNA overcharges the protein octamer by at least 74
negative charges (if not by $\approx 160$ charges since 86 charged
residues are buried inside the octamer). We thus expect a
considerable part of the terminal DNA to unwrap and to be part of
one or two tails. In Fig.~2 this corresponds to the wrapped chain
structures with tail that are found for small values of $l_{P}/\mu
$, $l_{P}/\mu <R$, and large values of $L$, $L>l_{iso}$.

It would be interesting to redo the experiments with DNA pieces
that are longer than 147 basepairs. For sufficiently large salt
concentrations there might be then the occurrence of rosettes (if
there is no interference with the partial disintegration of the
octamer). For the other limit, it might be appropriate to use the
argument for highly charged chains and spheres as given at the end
of the previous section. The linear charge density of DNA is very
high (2 phosphate groups per base pair, i.e., per 3.4\AA ).
Manning theory~\cite{manning78} indeed predicts that counterion
condensation reduces the linear charge density to $-e/l_{B}$ with
$l_{B}=7$\AA . Also the charge of the histone octamer is so high
that counterion condensation is important. For that case we
predicted above the unwrapping to take place around $l_{P}=\Omega
R^{2}/b$. However, since the DNA persistence length is smaller
than $R^{2}/b\approx \left( 50\,\text{\AA }\right) ^{2}/1.7$\AA\
$\approx 1500 $\AA\ we expect the wrapped state to be stable at
low ionic strength and no rosettes should occur in this limit.

Concluding, we have presented a scaling theory for the
complexation between a charged chain and an oppositely charged
sphere for the two limits of high and low ionic strength. For
strong screening one encounters with increasing chain stiffness a
strongly discontinuous transition from a wrapped to an open
multi-leafed rosette structure as already predicted in Ref.~\cite
{schiessel00}. In the case of weak screening one has again wrapped
structures for sufficiently flexible chains and rosettes for
stiffer chains. The unwrapping transition, however, is rather
smooth in this case. That this is a characteristics of a system
with long-ranged interaction has been further supported by
extending this theory to power law attraction with arbitrary
exponents and to the nonlinear limit of highly charged systems. We
hope that the presented ''phase diagrams'' for the two cases are
helpful for further theoretical and experimental studies.

\acknowledgements I am grateful to Anna Akinchina and Per Linse
for sending their paper, Ref.~\cite{akinchina02}, prior to
publication, and for helpful comments.

\end{document}